\documentstyle[11pt,aasms4,epsf,epsfig]{article}
\def\la{\lower0.6ex\vbox{\hbox{$ \buildrel{\textstyle 
<}\over{\sim}\ $}}}
\def\ga{\lower0.6ex\vbox{\hbox{$ \buildrel{\textstyle
>}\over{\sim}\ $}}}

\def\beq{\begin{equation}}
\def\eeq{\end{equation}}
\def\beginapjbib{\begingroup \section*{\large \bf References}
         \parskip=.5ex plus 1.0pt
         \def\bibitem{\par \noindent \hangindent\parindent
                \hangafter=1}}
\def\endapjbib{\par \endgroup}
\def\alwaysmath#1{{\ifmmode{#1}\else{$#1$}\fi}}
\def\he#1{\hbox{\alwaysmath{{}^{#1}}{\rm He}}}

\def\hii{H\thinspace{$\scriptstyle{\rm II}$}~}

\def\etal{{\it et al.}~}
\def\ie{{\it i.e.},~}
\def\eg{{\it e.g.},~}
\def\3he{$^3$He}
\def\4he{$^4$He}
\def\6li{$^6$Li}
\def\7li{$^7$Li}
\def\3h{$^3$H}
\def\Yp{Y$_{\rm P}$~}

\begin{document}

\vskip 0.7in
 
\begin{center} 

{\Large{\bf THE EVOLUTION OF HELIUM AND HYDROGEN IONIZATION CORRECTIONS 
AS \hii REGIONS AGE}}
 
\vskip 0.4in
{Ruth Gruenwald$^1$, Gary Steigman$^2$, and Sueli M. Viegas$^1$}
 
\vskip 0.2in
{\it $^1${Instituto Astron$\hat{o}$mico e Geof\' \i sico, 
Universidade de S$\tilde{a}$o Paulo, \\
S$\tilde{a}$o Paulo, S.P. 04301-904, BRASIL}}\\

\vskip 0.1in 
{\it $^2${Departments of Physics and Astronomy,
The Ohio State University, \\ 
Columbus, OH 43210, USA}}\\
\newpage

{\bf Abstract}
\end{center}

Helium and hydrogen recombination lines observed in low-metallicity,
extragalactic, \hii regions provide the data used to infer the primordial
helium mass fraction, Y$_{\rm P}$.  In deriving abundances from observations,
the correction for unseen neutral helium or hydrogen is usually assumed
to be absent; \ie the ionization correction factor is taken to be unity
($icf \equiv 1$). In a previous paper (VGS), we revisited the question of
the $icf$ for \hii regions ionized by clusters of young, hot, metal-poor
stars, confirming earlier work which had demonstrated a ``reverse'' ionization
correction: $icf < 1$.  In VGS the $icf$ was calculated using more nearly
realistic models of inhomogeneous \hii regions, revealing that for those
\hii regions ionized by young stars with ``hard'' radiation spectra the
$icf$ is reduced even further below unity compared to homogeneous models.
Based on these results, VGS suggested that the published values of \Yp
needed to be reduced by an amount of order 0.003.  As star clusters age,
their stellar spectra evolve and so, too, will their $icf$s.  Here the
evolution of the $icf$ is studied, along with that of two, alternate,
measures of the ``hardness" of the radiation spectrum.  The differences
between the $icf$ for radiation-bounded and matter-bounded models are also
explored, along with the effect on the $icf$ of the He/H ratio (since He
and H compete for some of the same ionizing photons).  Particular attention
is paid to the amount of doubly-ionized helium predicted, leading us to
suggest that observations of, or bounds to, He$^{++}$ may help to discriminate
among models of \hii regions ionized by starbursts of different ages and
spectra.  We apply our analysis to the Izotov \& Thuan (IT) data set utilizing
the radiation softness parameter, the [OIII]/[OI] ratio, and the presence
or absence of He$^{++}$ to find $0.95~\la icf~\la 0.99$.  This suggests
that the IT estimate of the primordial helium abundance should be {\it
reduced} by $\Delta $Y$ \approx 0.006 \pm 0.002$, from $0.244 \pm 0.002$
to $0.238 \pm 0.003$.

\newpage

\noindent

\section{Introduction}

As the second most abundant element in the universe, whose abundance even today
is dominated by its early universe production during Big Bang Nucleosynthesis
(BBN), \he4 plays a key role in testing the consistency of the standard, hot,
big bang model of cosmology and in using primordial nucleosynthesis as a probe
of cosmology and particle physics (for a recent review and further references,
see Olive, Steigman \& Walker 2000).  To minimize the unavoidable contamination
from post-BBN production of \he4, low-metallicity, extragalactic \hii regions
are the targets of choice for deriving the primordial abundance.  Observations
of large numbers of such \hii regions have led to estimates of the primordial
helium mass fraction, $Y_{\rm P}$, whose {\it statistical} uncertainties are
very small, $\approx$ 1\% (\eg see, Olive \& Steigman 1995 (OS), Olive, Skillman,
\& Steigman 1997 (OSS), Izotov, Thuan \& Lipovetsky 1994, 1997 (ITL), Izotov
\& Thuan 1998 (IT), Peimbert, Peimbert, \& Ruiz 2000 (PPR)).  Now, more than
ever before, it is crucial to attack potential sources of {\it systematic}
error.  In a previous paper (Viegas, Gruenwald, \& Steigman 2000 (VGS)) we
explored the size of the ionization correction for unseen neutral hydrogen
and/or helium in \hii regions ionized by young, hot, metal-poor stars. In
VGS models of more nearly realistic, inhomogeneous \hii region models were
compared with the data set assembled by IT, leading to the conclusion that,
on average, $icf < 1$ for the IT \hii regions.  Since IT (as well as almost
all other analyses of \hii region observations) {\it assumed} $icf \equiv 1$,
VGS suggested that their derived value of $Y_{\rm P}$ needed to be reduced
by $\approx 0.003$, from $Y_{\rm P} \approx 0.244$ to $Y_{\rm P} \approx
0.241$.  Some support for this lower value comes from the recent observations
by PPR of a relatively metal-rich SMC \hii region for which they derive
Y(SMC)$ = 0.2405 \pm 0.0017$, as well as from an independent theoretical
study by Sauer \& Jedamzik (2001; hereafter SJ).

In VGS models were calculated of \hii regions ionized by the radiation from
starbursts of two different ages (Cid-Fernandes \etal 1992): $t = 0$ and $t
= 2.5$~Myr.  The predictions for these two cases are in stark contrast, with
$icf < 1$ for the young star cluster and $icf > 1$ for the older one.  The
difference is easy to understand as a consequence of the evolving radiation
spectrum.  The young star cluster has many massive, hot stars which produce
a ``hard" spectrum with many helium-ionizing photons.  As the star cluster
ages, the more massive stars die, and the spectrum softens, reducing the
relative number of helium to hydrogen ionizing photons.  However, the evolution
of the radiation spectrum from a cluster of stars is not monotonic.  As time
goes by, stellar winds strip away the atmospheres of lower mass stars exposing
their hotter interiors and the resulting stellar spectra harden, producing
relatively more He$^{0}$ and He$^{+}$ ionizing photons. Eventually, even these
stars end their lives and the spectrum softens, the overall flux of ionizing
radiation decreases, and the \hii region fades away.

When comparing with the IT data, in order to distinguish between \hii regions
ionized by ``hard" spectra for which $icf < 1$ and those ionized by ``soft"
spectra leading to $icf > 1$, VGS utilized the Vilchez-Pagel ``radiation
softness parameter" (Vilchez \& Pagel 1998)
\beq
\eta \equiv ({n(0^+) \over n(S^+)})({n(S^{++}) \over n(O^{++})}),
\label{eta}
\eeq
finding that the IT data prefer a hard spectrum (log~$\eta~\la 0.4$),
favoring a ``reverse" ionization correction ($icf < 1$).

In this paper the evolution of model \hii regions is studied tracking
the changes in the $icf$, the radiation softness parameter $\eta$, the
abundance of doubly ionized helium (He$^{++}/$H$^{+}$), and an alternate
radiation softness parameter proposed by Ballantyne, Ferland, \& Martin
(2000; hereafter BFM), the ratio [OIII]$\lambda$5007/[OI]$\lambda$6300.
In \S2 we follow the evolution of the radiation spectrum using the
starburst models of Cid-Fernandes \etal (1992), concentrating on the
{\it relative} changes in the hydrogen and helium ionizing fluxes and
their implications for the evolution of the $icf$, $\eta$, [OIII]/[OI],
and the HeII(4686)/H${\beta}$ ratio.  Since hydrogen and helium compete
for the same ionizing photons, the sizes of the H$^{0}$ and He$^{0}$
zones may depend on the ``true" He/H ratio.  This effect is explored
in \S3.  In \S4 we model \hii regions ionized by the composite spectra
of starbursts of different ages, and investigate the observational
consequences of superposed \hii regions (see, also, VGS).  Since it
is the structure of the ``outer" zones in an \hii region which most
influences the $icf$, in \S5 the predictions of radiation-bounded
and matter-bounded models are compared.  Our results are discussed
in \S6 and our conclusions presented in \S7.

\section{Evolution of Starburst Spectra And The $icf$}

\begin{figure}[ht]
	\centering
	\epsfysize=3.78truein
\epsfbox{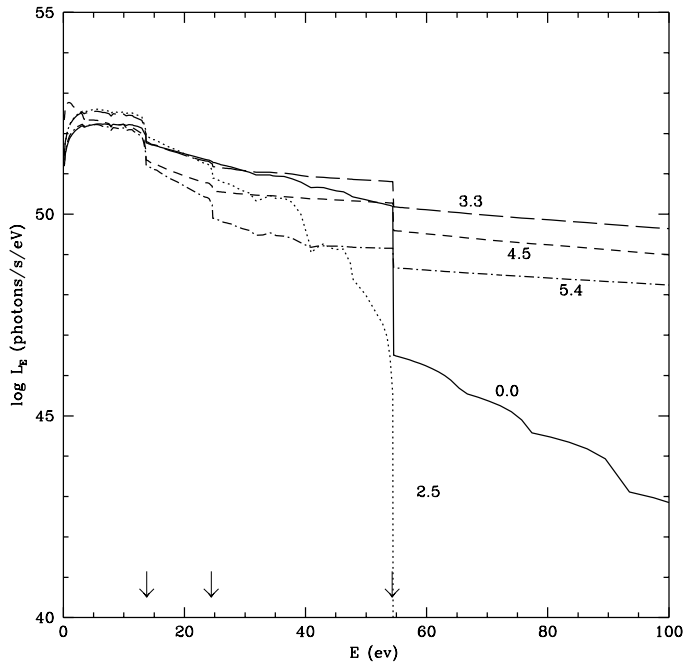}
	\caption{\small{The spectral distribution (photons~s$^{-1}$eV$^{-1}$)
	of the radiation from a $10^{6}M_{\odot}$ starburst at five different 
	epochs in its evolution: $t = 0$ (solid), 2.5 Myr (dotted), 3.3 Myr 
	(long-dashed), 4.5 Myr (short-dashed), and 5.4 Myr (dot-dashed).  The
        breaks in the spectra occur at the hydrogen and helium ionization
	edges, indicated by the arrows.}}
	\label{spectra}
\end{figure}

\begin{figure}[ht]
	\centering
	\epsfysize=3.78truein
\epsfbox{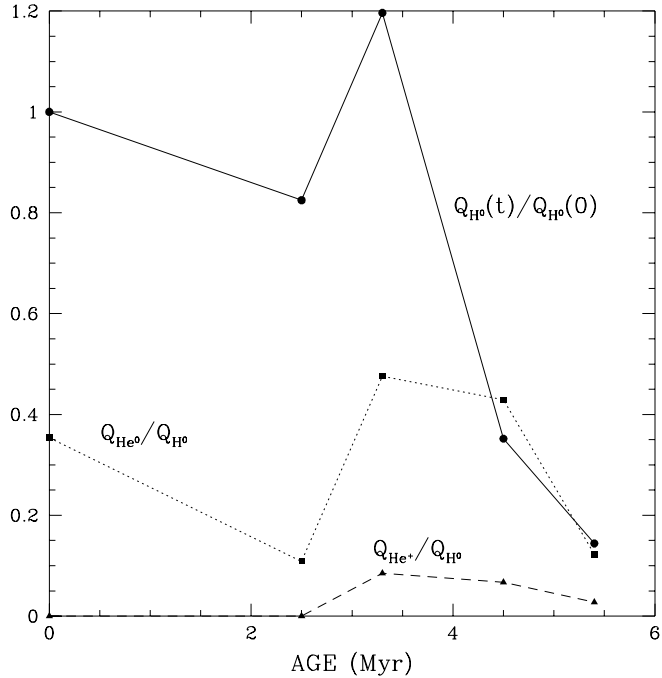}
	\caption{\small{The evolution of the ionizing fluxes 
as the starburst ages.  The solid line traces the evolution
of hydrogen-ionizing photons ($E \geq 13.6$~eV).  The dotted
line follows the ratio of all He-ionizing ($E \geq 24.6$~eV) 
to H-ionizing fluxes, while the dashed line is the same for
the He$^{+}$-ionizing photons ($E \geq 54.4$~eV) alone.}}  
	\label{Qs}
\end{figure}

Cid-Fernandes \etal (1992) explored the time-evolution of the
radiation from starbursts of different total stellar masses.  
The models utilize the Maeder stellar evolution models and 
adopt a Salpeter IMF (slope = 1.35) with an upper mass limit
of 130~$M_{\odot}$.  In Figure~\ref{spectra} are shown the 
spectral flux distributions for a starburst of $10^{6}M_{\odot}$
at five different times in its evolution: $t =$~0.0, 2.5, 
3.3, 4.5, 5.4 Myr.  To more clearly expose those aspects of 
these spectra which are key to understanding the evolution 
of the $icf$, in Figure~\ref{Qs} are shown the time-evolution 
of various ratios of the hydrogen (H$^{0}$) and helium (He$^{0}$, 
He$^{+}$) ionizing fluxes.

Initially, due to the presence of massive, hot stars, relatively large
amounts of helium ionizing photons are present resulting, in general,
in $icf < 1$ (see VGS).  As the more massive stars evolve and die, the
overall flux from the starburst decreases with the relative number of
helium ionizing photons falling even faster, so that the $icf$ increases
and, by $t = 2.5$~Myr, $icf > 1$ (VGS). However, as the starburst continues
to age, stellar winds strip away the atmospheres of the massive stars, 
exposing their hotter interiors.  As a result, both the overall flux and 
the relative numbers of helium ionizing photons increase.  In these more 
evolved \hii regions ($t \sim 3$~Myr) the flux of He$^{+}$-ionizing photons
is significant and detectable amounts of doubly-ionized helium should be 
present.  As the star cluster continues to evolve, the overall flux of 
ionizing radiation decreases and so, too, does the ratio of helium-ionizing 
to hydrogen-ionizing photons.  From Fig.~\ref{Qs} it can be anticipated
that while a $~\ga 5$~Myr old \hii region will have an $icf > 1$, similar 
to that of a $\sim 2.5$~Myr starburst, detectable amounts of He$^{++}$ may 
be present in the former while absent in the latter, perhaps permitting an 
observational discrimination between the two.

\subsection{Photoionization Models} 

Here, as in VGS, the AANGABA photoionization code (Gruenwald \&
Viegas 1992) is used to model the gas whose distribution is taken
to be spherically symmetric and homogeneous.  Since we are specifically
interested in the helium -- hydrogen $icf$ for low-metallicity \hii
regions, a metal-poor chemical composition (0.1 solar) is chosen.
Initially He/H = 0.083 (Y $\approx 0.25$) is adopted, but later 
(\S3) the consequences for the $icf$ of different helium abundances 
are explored.  The radiation intensity is characterized by the number 
of ionizing photons above the Lyman limit, Q$_H$, which is directly 
related to the mass and the IMF of the stellar cluster.  Along with 
the gas density, $n$, Q$_H$ defines the model \hii regions.  The
popular ionization parameter, U, is proportional to Q$_H$/$n$, so 
that a grid of models at fixed density, characterized by different 
choices of U and R$_i$ (the inner radius of the model \hii region), 
corresponds to the identical grid labelled by {\it different} values
of Q$_H$.

In VGS the $icf$ (called ``ICF" in that paper) was defined by relating
the total He/H ratio to that of He$^{+}/$H$^{+}$ alone.  As a result,
any He$^{++}$ was included along with unseen neutral H and He.  However,
lines from the recombination of He$^{++}$ are observed in many \hii
regions, so that a more generally applicable definition, adopted in
this paper, is
\beq
{{\rm He} \over {\rm H}} \equiv icf{({\rm He}^{+} + {\rm He}^{++}) \over
{\rm H}^{+}}.\label{icf}
\eeq
For \hii regions with He$^{++} \ll $~He$^{+}$, $icf \approx$~ICF, while in
the presence of small, but observable amounts of He$^{++}$, $icf ~\la$~ICF.

\begin{figure}[ht]
	\centering
	\epsfysize=3.78truein
\epsfbox{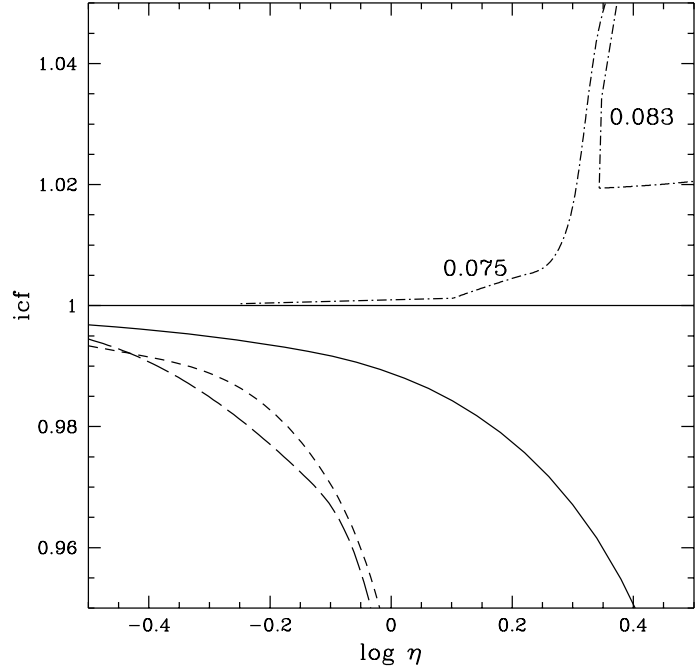}
	\caption{\small{The $icf$ (see eq.~\ref{icf}) versus the
	radiation softness parameter log~$\eta$ relations for
	\hii regions of all ages ($0.0 \leq t \leq 5.4$~Myr).
	The solid line is for $t = 0$, the long-dashed line is
	for 3.3 Myr, the short-dashed line is for 4.5 Myr, and
	the dot-dashed lines are for 5.4 Myr.  The 2.5 Myr model,
	for which log~$\eta ~\ga 0.8$, is off scale.  The two
        curves for the 5.4 Myr starburst are for two choices
        of the helium abundance, He/H = 0.075 and 0.083 (see
        \S\ref{he/h}).}}
	\label{icfvseta}
\end{figure}

Using AANGABA, grids of model \hii regions ionized by the radiation from
starbursts of differing masses (\ie different Q$_{H}$), were created at
each of the cluster ages identified above.  In Figure~\ref{icfvseta} the
results are shown in the $icf$ -- log~$\eta$ plane.  As anticipated from
Figures \ref{spectra} and \ref{Qs}, $icf < 1$ for the hard spectra of the
zero-age, 3.3, and 4.5 Myr starbursts, while $icf > 1$ for the soft spectra
of the 2.5 and 5.4 Myr models.  Note that the 2.5 Myr starburst is offscale
in Fig.~\ref{icfvseta} (where log $\eta~\le 0.5)$ since for the very soft
spectrum of this starburst, log $\eta ~\ga 0.8$ (see VGS and Fig.~\ref
{etao3o1}); for the IT data, log $\eta ~\la 0.4$.  For the 5.4 Myr
starburst the $icf - \eta$ relation depends on the helium abundance as
discussed below in \S\ref{he/h}.  For the $t = 0$, 3.3, 4.5 Myr models,
and the 5.4 Myr model with He/H = 0.075, as the intensity of the starburst
increases, $\eta$ decreases and $icf \rightarrow 1$.  In contrast, for the
5.4 Myr model with He/H = 0.083, as the intensity of the starburst increases,
at first the $icf$ {\it decreases} at nearly constant $\eta$, then $\eta$
{\it increases} (to log~$\eta \sim 0.6$) at nearly constant $icf$, and
finally, as the size of the starburst is increased even further, $\eta$
{\it decreases} (to log~$\eta \sim 0.4$) as $icf \rightarrow 1$.

\begin{figure}[ht]
	\centering
	\epsfysize=3.78truein
\epsfbox{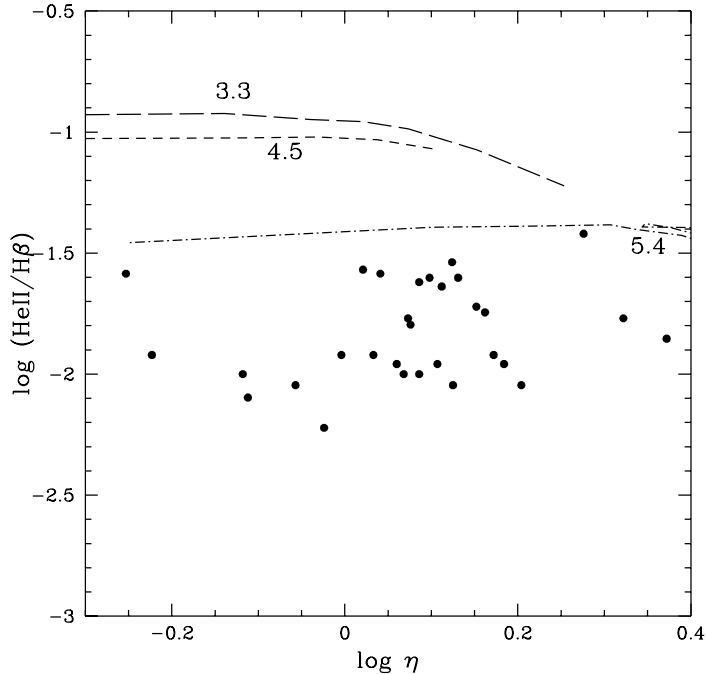}
	\caption{\small{The HeII/H$\beta$ ratio versus log~$\eta$
	for \hii regions of differing ages.  The starburst ages are
	labelled on the curves.  There are two dot-dashed curves
        for the 5.4 Myr starburst, corresponding to the two helium
        abundances (see Fig.~\ref{icfvseta}).  Note that for models
        of zero age and 2.5 Myr old starbursts, the predicted HeII
        emission is so small as to be off scale.  The filled circles
        are the IT data.
	}}
	\label{he2vseta}
\end{figure}

We note that our models compute the line intensities integrated
over the whole \hii region while the IT data to which we will compare is
taken using a $2'' \times 300''$ slit.  Since the slit is likely oriented
to include the largest possible fraction of the \hii region, it is
reasonable to compare our models with their observations.  However, since
the $icf$ is determined by the outermost zone of the \hii region, for those
radiation-bounded regions which may be larger than the observing slit, the
deviation of the $icf$ from unity will be reduced relative to our results.
For example, for long slit observations of \hii regions smaller than the
slit length but larger than the slit width, $|1 - icf|$ can differ by up
to 50\% from our model predictions.

In Figure~\ref{he2vseta} the model-predicted HeII(4686)/H$\beta$ flux
ratios are shown for starbursts of different ages, as a function of the
radiation softness parameter $\eta$.  Note that no curves appear for
the zero-age and 2.5 Myr models because they predict that no detectable
He$^{++}$ should be present (at a level HeII/H$\beta ~\ga 10^{-3}$).
Also shown are the relevant data from IT who have detected He$^{++}$
in $\sim 30$ \hii regions.  It is noteworthy that {\it more} HeII(4686)
is observed than is predicted for the young \hii regions, while the
opposite appears to be the case for the older \hii regions.  This strongly
suggests that ``real" \hii regions may be ionized by a mixture of starbursts
of different ages or, that what is observed could be the superposition
of several \hii regions within the telescope aperture (see VGS).  If so,
then the ``true" $icf$ must reflect some appropriately weighted average
of the various $icf$s shown in Figure~\ref{icfvseta}.  We return to this
issue in \S\ref{composite}, but note here that for the starburst spectrum
employed by SJ, it is unlikely that any detectable He$^{++}$ will be present.
Clearly, the presence or absence of He$^{++}$ can be a valuable clue to the
hardness of the ionizing radiation spectrum.  Before pursuing these issues,
we first consider the alternative radiation softness parameter proposed by BFM.

\subsection{An Alternate Radiation Softness Parameter?}

There are advantages and disadvantages to using the Vilchez-Pagel 
radiation softness parameter $\eta$ (see eq.~\ref{eta}).  Since
$\eta$ requires the {\it ionic} ratios, it may be more convenient
to employ a related parameter, similar to one introduced by Skillman
(1989), which is simply a ratio of {\it line} ratios,
\beq
\eta' \equiv ({I(\lambda 3727) \over I(\lambda 4959+\lambda 5007)})({I(\lambda 6312)
\over I(\lambda 6717+\lambda 6731)}).
\eeq
From our models we find that for the IT data range of log~$\eta < 0.4$,
$\eta'$ and $\eta$ are closely related by,
\beq
{\rm log}~\eta' = -1.82 + 0.88~{\rm log}~\eta.
\eeq
In our attempt to compare this with the IT {\it line} and {\it ionic} ratio
data we have uncovered several inconsistencies between their line intensity
ratios and their ionic ratios, which contribute to a spread around a best
fit relation from the data of
\beq
{\rm log}~\eta'(IT) = -1.85 + 0.76~{\rm log}~\eta.
\eeq
To facilitate comparison with previous analyses, in this paper we choose
to utilize the original Vilchez-Pagel parameter $\eta$.

For the
relatively hard spectra corresponding to starbursts of $t =$ 0.0, 
3.3, and 4.5 Myr, $\eta$ decreases monotonically as the intensity
of the starburst, Q$_{H}$, increases (\ie $\eta$ and Q$_{H}$ are
anticorrelated).  For these models there is also an anticorrelation
between the $icf$ (which is $ < 1$) and $\eta$ (see Fig.~\ref{icfvseta}).
In contrast, the $\eta$ -- Q$_{H}$ and $\eta$ -- $icf$ relations
are {\it not} monotonic for the softer spectra from starbursts of
2.5 and 5.4 Myr (for which $icf > 1$).  Furthermore, for the latter
starburst the character of the $icf - \eta$ relation changes with
helium abundance.

\begin{figure}[ht]
	\centering
	\epsfysize=3.78truein
\epsfbox{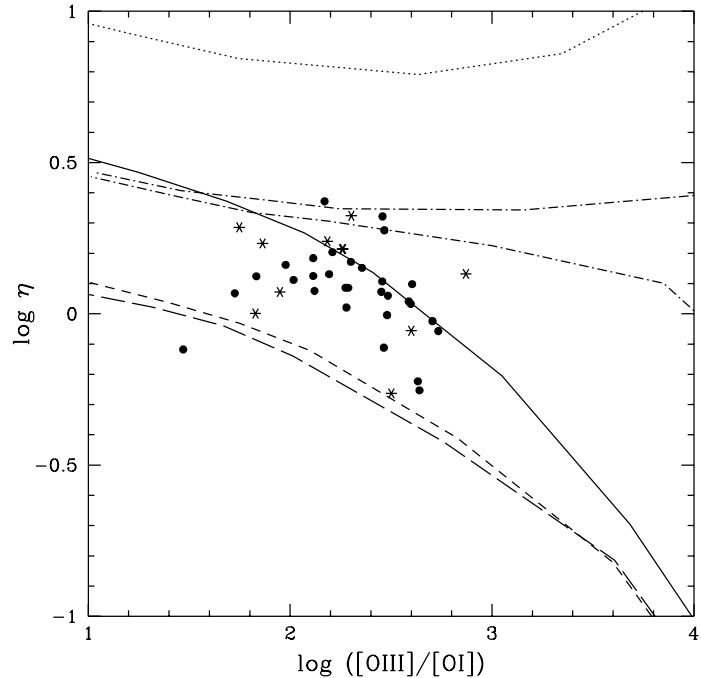}
	\caption{\small{The radiation softness parameter $\eta$ versus
        the [OIII]/[OI] ratio for \hii regions of all ages (the line
        types are as in Figs.~\ref{spectra} and \ref{icfvseta}).  Note
        that there are two curves (dot-dashed) for the 5.4 Myr models,
        corresponding to He/H = 0.075 (lower of the two) and 0.083 (see
        \S3).  The filled circles are the IT \hii regions with observed
        He$^{++}$; the stars are for those regions without He$^{++}$.}}
	\label{etao3o1}
\end{figure}

BFM proposed an alternate radiation softness parameter, the ratio of
[OIII]$\lambda5007$ to [OI]$\lambda6300$.  This ratio of observed fluxes
is independent of the \hii region metallicity.   We have tracked this
ratio in our models and in Figure~\ref{etao3o1} the relationship
between the Vilchez-Pagel radiation softness parameter, $\eta$, and the
BFM parameter [OIII]/[OI], is shown for a series of \hii region models
ionized by starbursts of different ages.  The variable along the curves
is the intensity of the starburst, proportional to Q$_{H}$.  For all
starbursts as Q$_{H}$ increases so, too, does the [OIII]/[OI] ratio.
Figure~\ref{etao3o1} exposes a crucial difference between $\eta$ and
[OIII]/[OI]: whereas all spectra, whether ``hard" or ``soft", span the
{\it entire} range in [OIII]/[OI], the ranges covered by $\eta$ are
{\it different} for hard and soft spectra.  While $\eta$ may not be
the ideal radiation softness parameter, the [OIII]/[OI] ratio has
virtually no sensitivity to the spectral shape; rather, [OIII]/[OI]
provides a measure of the {\it intensity} of the ionizing radiation
field.

There is another problem if the [OIII]/[OI] ratio is used to provide an
estimate of the hardness of the radiation spectrum.  The [OI]$\lambda6300$
emission-line is produced in the recombination region, so its intensity
is very dependent on the optical depth at the Lyman limit, $\tau_{LL}$.  
We return to this point in \S\ref{mbrb}, but simply remark here that for 
most models, the [OIII]/[OI] ratio decreases by more than 3 orders of 
magnitude as $\tau_{LL}$ increases from of order unity (matter-bounded) 
to of order 10$^4$ (radiation-bounded).  In contrast, although the O$^+$
and S$^+$ fractions are each sensitive to $\tau_{LL}$, $\eta$ varies much 
less with $\tau_{LL}$ since it involves their {\it ratio} O$^+$/S$^+$.  

\begin{figure}[ht]
	\centering
	\epsfysize=3.78truein
\epsfbox{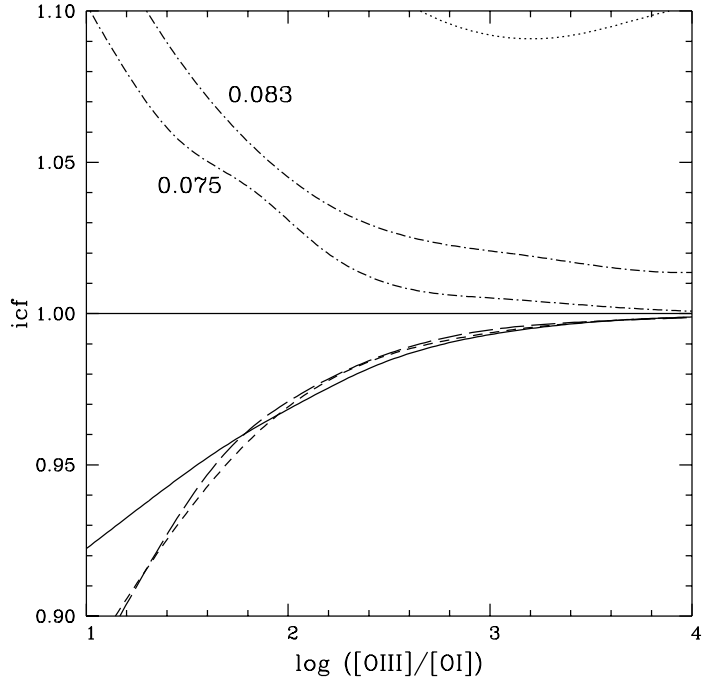}
	\caption{\small{The $icf$ versus the [OIII]/[OI] ratio for
	\hii regions of all ages.  The line types are as in Figs.
	\ref{spectra}, \ref{icfvseta}, and \ref{etao3o1}.  The labels
        on the dot-dashed curves (5.4 Myr starburst) are the values
        of He/H.}}
	\label{icfo3o1}
\end{figure}

The relevant IT data are also shown on Fig.~\ref{etao3o1}.  Notice that
the IT \hii regions, with $-0.3 ~\la$ log~$\eta~\la +0.4$, lie between
the models for starbursts of all ages, but much below that for a 2.5 Myr
starburst for which log~$\eta~\ga 0.8$.  This suggests (see VGS) that 
starbursts of $\approx 2.5$~Myr, with $icf > 1$, are not dominant in the
\hii regions observed by IT.  The $icf$s for starbursts of differing ages
are shown as a function of the [OIII]/[OI] ratio in Figure~\ref{icfo3o1}.
Once again, notice that a measurement of the [OIII]/[OI] ratio does {\it
not} determine either the size, or the sign, of the $icf$.

\begin{figure}[ht]
	\centering
	\epsfysize=3.78truein
\epsfbox{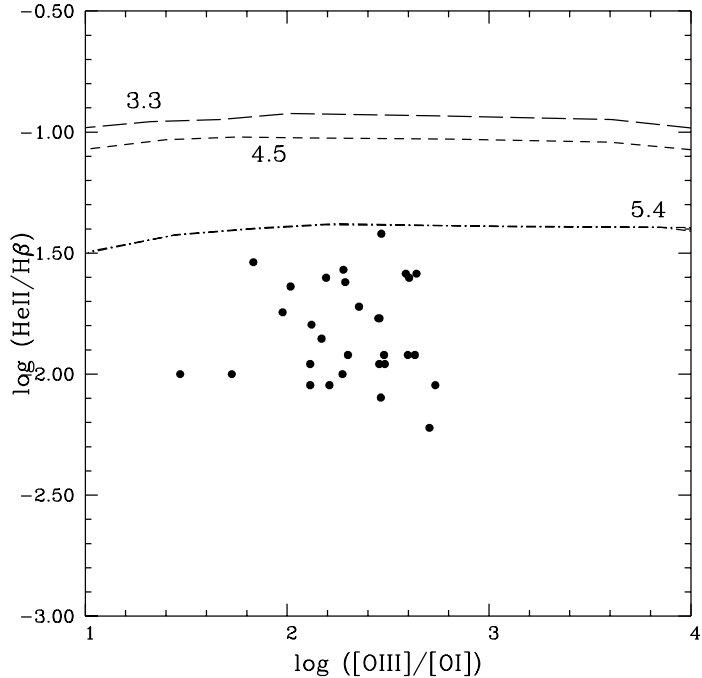}
	\caption{\small{The HeII/H$\beta$ ratio versus [OIII]/[OI]
	for \hii regions of differing ages.  The starburst ages are
	labelled on the curves.  Note that for models of zero age
        and 2.5 Myr starbursts the predicted HeII emission is so
        small as to be off scale.  The filled circles are from IT.}}
	\label{he2vso3o1}
\end{figure}

The predicted HeII/H$\beta$ ratios are shown as a function of [OIII]/[OI]
for starburst models of different ages in Figure~\ref{he2vso3o1}.  As was
already seen in Fig.~\ref{he2vseta}, no significant HeII is present in the
young starbursts ($t ~\la 2.5$~Myr), while higher than observed HeII/H$\beta$
ratios are predicted for the older \hii regions.

Although we are in agreement with BFM that the $icf \rightarrow 1.0$ as
[OIII]/[OI] increases, we emphasize that for the [OIII]/[OI] range of the IT
data ($1.4~\la$~log([OIII]/[OI])~$\la 2.9$), there are significant ionization
corrections for starbursts of all ages.  This is in contrast to the BFM
suggestion that $icf = 1$ when [OIII]/[OI] $\ga 300$.  At the same time,
the Pagel \etal (1992) proposal that $icf = 1$ for log~$\eta < 0.9$ is
unsupported by either our results (see Fig.~\ref{icfvseta}) or those of SJ.
For the $\eta$ and [OIII]/[OI] ranges of the IT data, significant ionization
corrections (at the few percent level) -- most of them with $icf < 1$ -- are
to be expected (see, also, SJ).

\section{The Effect Of The Helium Abundance On The Ionization Structure}
\label{he/h}

Although models of the low metallicity \hii regions targeted for
determining the primordial helium abundance almost always adopt 
low heavy element abundances, little attention is usually given
to the assumed helium abundance.  For example, the helium abundance
(by number) is He/H = 0.10 in Stasinska's (1990) models which were 
used by Pagel \etal 1992 to analyze their observed, low-metallicity
\hii regions.  This helium abundance, Y~$\approx 0.29$, is even
higher than the solar value (Bahcall, Pinsonneault, \& Basu 2001).
The Armour \etal (1999) and BFM analyses use the Orion abundance,
He/H = 0.095 (Y~$\approx 0.28$).  VGS set Z = Z$_\odot$/10 and chose
He/H = 0.083 (Y~$\approx 0.25$).

Since helium competes with hydrogen for some of the same ionizing 
photons, the precise value of the helium abundance may affect the
ionization structure of the \hii region (Gruenwald \& Viegas 2000).
As we are tracking small deviations in the derived helium abundance,
it is important to explore the effect of the adopted helium abundance
on the model predictions of the $icf$, along with the other diagnostics
such as $\eta$ and the [OIII]/[OI] ratio.  For starbursts of all ages
model \hii regions were computed for helium abundances varying from
He/H = 0.075 (Y~$\approx 0.23$) to 0.10 (Y~$\approx 0.29$).  From
these models the following picture emerges.

The \hii regions formed by massive starbursts (high Q$_H$) are large,
with very thin transition zones, leading to an $icf \rightarrow 1$.
For the more realistic starbursts (smaller Q$_H$), the He$^0$ and
H$^0$ zones are larger and the $icf$ deviates from unity, $> 1$ or
$< 1$, depending on the hardness of the radiation spectrum.  With the
exception of the 5.4 Myr starburst, there are no significant variations
of the $icf$ (or of the other diagnostics) with He/H.

\begin{figure}[ht]
	\centering
	\epsfysize=3.78truein
\epsfbox{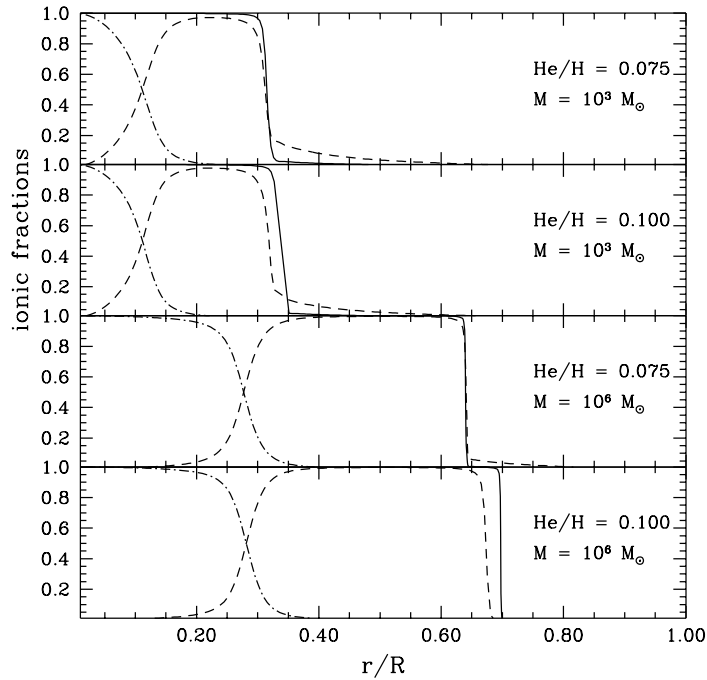}
	\caption{\small{The radial distributions of the H$^{+}$
        (solid), He$^{+}$ (dashed), and He$^{++}$ (dot-dashed)
        fractions.  $R$ is the radius at which the H$^{+}$ fraction
        has dropped to $10^{-4}$.  The two upper panels are for
        a $10^{3}M_{\odot}$ starburst and the two lower panels
        are for a $10^{6}M_{\odot}$ starburst.  In each pair
        of panels, the upper panel is for He/H = 0.075 and the
        lower one for He/H = 0.100.}}
	\label{panel}
\end{figure}

For the softer spectra of the oldest starbursts ($\ga 5$~Myr) there
are relatively fewer helium ionizing photons (see Fig.~\ref{Qs}) and
the corresponding \hii regions have some neutral helium inside the
ionized hydrogen zone so that $icf > 1$ (see Figs.~\ref{icfvseta} \&
\ref{icfo3o1}).  At fixed He/H, as the size of the starburst is reduced
(smaller Q$_{H}$) the transition zone thickens and the {\it relative}
contribution of He$^{0}$ increases, leading to an increase in the $icf$.
At the same time, the [OIII]/[OI] ratio (which tracks Q$_{H}$) decreases,
leading to an anticorrelation between the $icf$ and [OIII]/[OI].  Now, at 
fixed Q$_{H}$, as the He/H ratio increases the [OIII]/[OI] ratio, which 
depends mainly on Q$_{H}$, is essentially unchanged, but the transition 
zone, where helium has become neutral while hydrogen is still ionized, 
grows in size since there is now more helium competing for the same 
ionizing photons.  This effect is shown in Figure~\ref{panel} where
the radial distributions of the H$^{+}$, He$^{+}$, and He$^{++}$ fractions
are shown for two choices of the size of the starburst ($M_{*} \approx
10^{3}M_{\odot}$ and $M_{*} \approx 10^{6}M_{\odot}$) and two choices
for the helium abundance (He/H = 0.075, 0.100; Y $\approx 0.23$, 0.29).
For the smaller starburst, log([OIII]/[OI]) $\approx 2$, while for the
larger one, log([OIII]/[OI]) $\approx 3$ for all values of He/H from 0.075
to 0.100.  As a result, for a 5.4 Myr starburst, when the helium abundance
increases the curves shift upwards in the $icf$ -- [OIII]/[OI] plane
(see Fig.~\ref{icfo3o1}).

A much more dramatic change in the character of the curves occurs
in the $icf$ -- $\eta$ plane.  For He/H $\la 0.083$, as Q$_{H}$
increases {\it both} $\eta$ and the $icf$ decrease: $\eta$ and the
$icf$ are positively correlated (while $\eta$ and [OIII]/[OI] are
slightly anticorrelated; see Fig.~\ref{etao3o1}).  In contrast, for
He/H $\ga 0.083$ the variations of $\eta$ and the $icf$ with He/H,
and with the size of the starburst, are non-monotonic. For these
higher helium abundances (Y $\ga 0.25$), not entirely relevant to
our study of low-metallicity \hii regions, the structure of the
transition zone is very sensitive to the precise helium abundance.
For example, for the He/H = 0.083 case shown in Fig.~\ref{icfvseta},
as Q$_{H}$ increases from smaller to larger starbursts, the $icf$
first decreases at nearly constant (slightly decreasing) $\eta$.
But, as Q$_{H}$ continues to increase, and before the $icf \rightarrow
1$, $\eta$ begins to increase at nearly constant $icf$.  Finally,
for values of Q$_{H}$ larger than that for any observed \hii regions
(\ie [OIII]/[OI] $\gg 10^{3}$), $\eta$ once again decreases and
so, too, does the $icf$ ($icf \rightarrow 1$).

\section{\hii Regions Ionized By Multiple Starbursts}
\label{composite}

\hii regions ionized by soft spectra will have $icf > 1$ and
relatively high values of $\eta$, while those ionized by harder
spectra will have $icf < 1$ and smaller $\eta$ values.  The IT
\hii regions have relatively small $\eta$ values ($-0.3 ~\la
$log~$\eta ~\la +0.4$; see Fig.~\ref{etao3o1}), confirming the
VGS and SJ conclusions that by neglecting the $icf$ (\ie by {\it
assuming} that the $icf = 1$) IT have likely overestimated the
primordial helium abundance.  The VGS and SJ results receive
further support from Figure~\ref{etao3o1} where the majority of
the IT data points are seen to lie between the zero-age model
and the $t = 3.3 - 4.5$ Myr starbursts which have harder spectra
and even smaller $icf$s.  However, the spread in the data highlights
the likelihood that at least some \hii regions may be ionized by
spectra from more than one starburst; there is support for this 
possibility from the Conti \& Vacca (1994) and the Conti, Leitherer, 
\& Vacca (1996) observations of substructure within \hii regions.  
Further support comes from the IT detection of doubly-ionized
helium in nearly 3/4 of their \hii regions at a level {\it above}
the zero-age prediction, but {\it below} the predictions of the 
3.3 -- 4.5 Myr models (see Figs.~4 \& 7).  Since models of all
three of these starbursts predict $icf < 1$, composite models
built from these spectra will have intermediate values of $\eta$,
and $icf < 1$.  However, what will happen if, to these spectra,
contributions are added from the 2.5 and 5.4 Myr starbursts?  
Clearly, the resultant \hii regions will have intermediate values 
of $\eta$, and an $icf$ which could either be $< 1$ or $> 1$.

We have explored two types of composite models to see when \hii regions
ionized by multiple starbursts will have sufficiently small values of
$\eta$ (and of [OIII]/[OI]), but an $icf$ which exceeds unity.   In VGS
we studied composite models of {\it separate} \hii regions (ionized by
$t = 0$ and 2.5 Myr starbursts) which may be included within the telescope
aperture, using various line ratios to exclude such contamination from
the IT data (see VGS for details).  Here we extend this analysis to
combinations of \hii regions ionized by $t = 0$ and 5.4 Myr starbursts;
starbursts of intermediate age all have $icf < 1$ and will only serve to
reduce the $icf$ from its $t = 0$ value.  However, before addressing such
models, first consider \hii regions ionized by the combined spectra of
more than one starburst.  These multiple starburst possibilities are
limitless, so we have chosen to investigate the most ``conservative"
options: combining a zero-age starburst with either a 2.5 Myr or a 5.4
Myr starburst.  Including the 3.3 and/or 4.5 Myr starbursts would only
further reduce $\eta$ and the $icf$, as well as increase the predicted
amount of He$^{++}$ (which may already be too high).

\begin{figure}[ht]
	\centering
	\epsfysize=3.78truein
\epsfbox{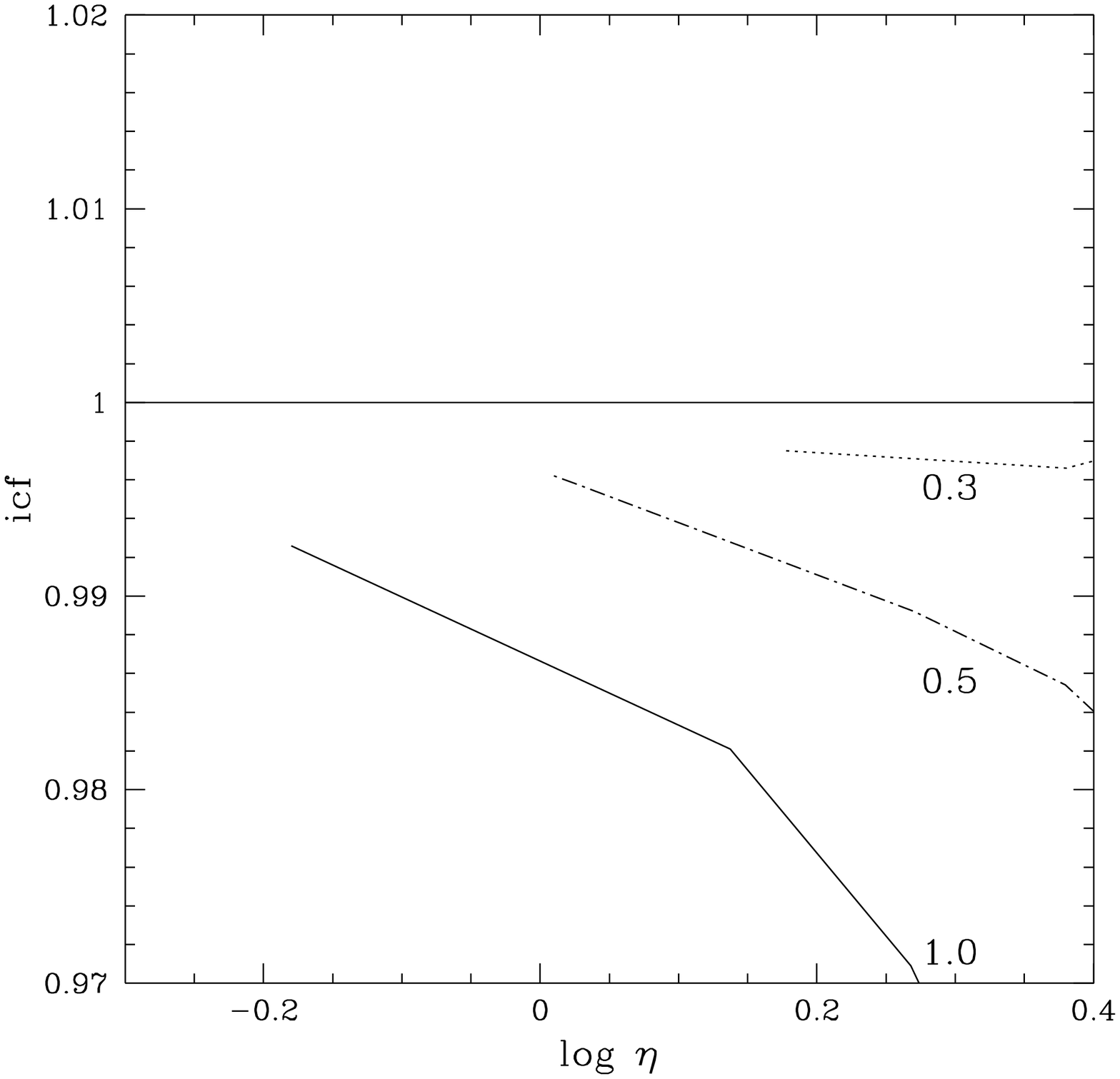}
	\caption{\small{The $icf - $log~$\eta$ relation for
        a series of \hii regions ionized by the composite
        spectra for zero-age and 2.5 Myr starbursts, subject
        to the constraint that [OIII]/[OI] $< 10^{3}$.  The
        numbers close to the curves are the fractions of the
        total flux coming from the zero-age starburst.}}
	\label{icfvseta25}
\end{figure}

Despite some similarities between the 2.5 Myr and 5.4 Myr spectra,
that of the 2.5 Myr starburst is ``softer", leading to \hii regions
with larger values of $\eta$ and of the $icf$.  The key question is,
what is the resulting $icf$ when the contribution from a zero-age
starburst (added to a 2.5 Myr starburst) is ``large enough" to have
reduced $\eta$ to the IT range (log~$\eta ~\la 0.4$)?  The results,
subject to the constraint [OIII]/[OI] $< 10^{3}$, are shown in
Figure~\ref{icfvseta25}.  It is the [OIII]/[OI] constraint that
sets the low $\eta$ limits and eliminates {\it any} composite
models with log~$\eta < 0.4$ and $icf > 1$, in agreement with
our previous conclusions in VGS.

\begin{figure}[ht]
	\centering
	\epsfysize=3.78truein
\epsfbox{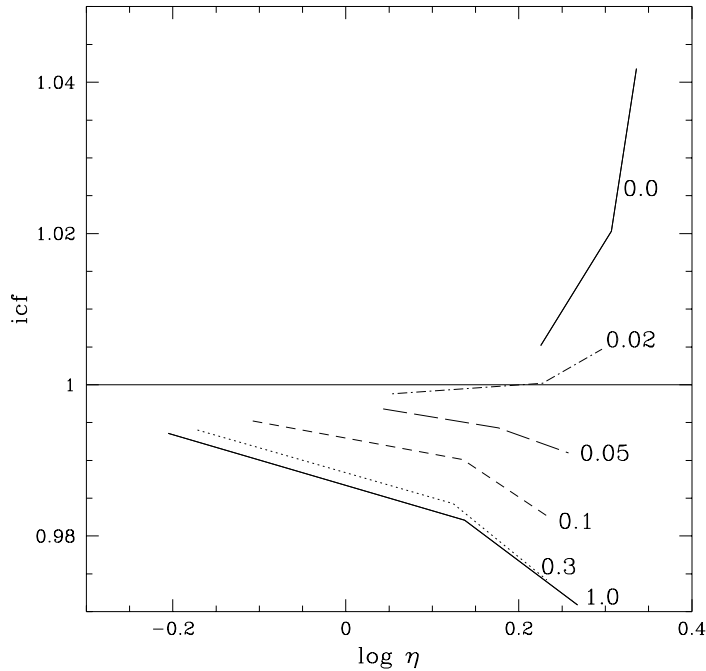}
	\caption{\small{The $icf - $log~$\eta$ relation for a
        series of \hii regions ionized by the composite spectra
        for zero-age and 5.4 Myr old starbursts, subject to the
        constraint that [OIII]/[OI] $< 10^{3}$.  For the 5.4 Myr
        starburst, He/H = 0.075 is adopted.  The numbers close
        to the curves are the fractions of the total flux coming
        from the zero-age starburst.}}
	\label{icfvseta54}
\end{figure}

\begin{figure}[ht]
	\centering
	\epsfysize=3.78truein
\epsfbox{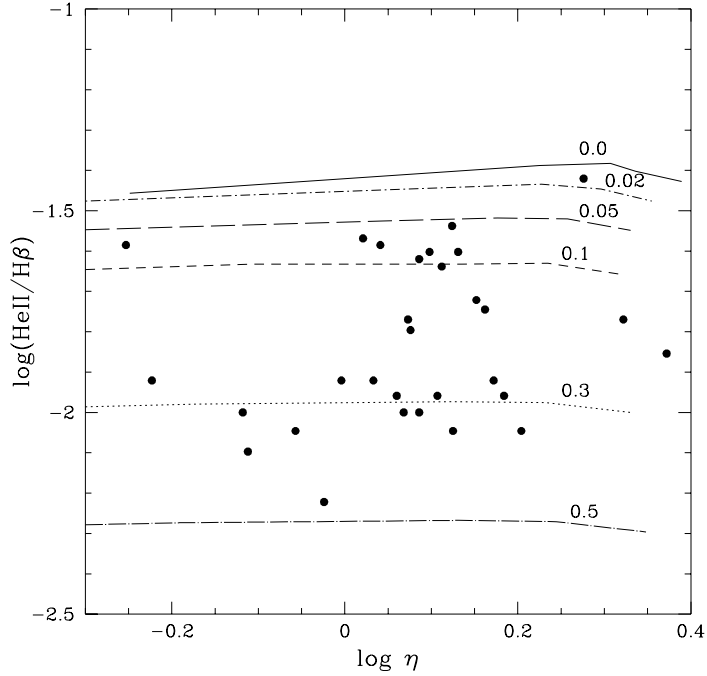}
	\caption{\small{The HeII/H$\beta - $log~$\eta$ relation
        for \hii regions with the same He/H, ionized by the 
        same composite spectra as in Fig.~\ref{icfvseta54} for 
        the zero-age and 5.4 Myr starbursts.  The filled circles 
        are the IT data.}} 
        \label{he2_75}
\end{figure}

The $icf$ versus $\eta$ relations for several combinations of the $t
= 0.0$ and $t = 5.4$~Myr models are shown in Figure~\ref{icfvseta54}.
Not until the contribution from the aging starburst is some 50 times
stronger than that from the infant starburst, does the $icf$ exceed
unity (and then, only for log~$\eta ~\ga 0.1$).  However, with such
a dominant 5.4 Myr old starburst, significant amounts of doubly-ionized
helium may be present.  The HeII/H$\beta$ predictions for these same
composite models are shown in Figure~\ref{he2_75}.  As anticipated,
those composite models sufficiently dominated by the 5.4 Myr starburst
to have $icf > 1$, will have more HeII than is observed in nearly all
of the IT \hii regions.  Consistent with our earlier results (VGS),
as well as those of SJ, it therefore seems unlikely that more than
a handful of the IT \hii regions can have $icf > 1$.

In agreement with SJ, we note from our exploration of these composite models
that the dividing line between $icf < 1$ and $> 1$ corresponds to a ratio
of helium-ionizing to hydrogen-ionizing fluxes of $\approx 0.15$.  Since
our spectra for the individual as well as the composite starbursts differ
in detail from those of SJ, this result provides supporting evidence that
this ratio, and not the detailed shape of the ionizing spectra, is key to
determining the sign of $icf - 1$.

For {\it separate} \hii regions, ionized by $t = 0$ and 5.4 Myr starbursts,
which may be included within the telescope aperture, the situation is less
straightforward.  In our exploration of such composite models we have found
it possible that the combined \hii region can have an $icf > 1$ while satifying
log~$\eta ~\la 0.4$ along with the [OIII]/[OI] and HeII/H$\beta$ constraints.
Such models always have log~$\eta ~\ga 0.25$.  A glance at Figure~\ref{etao3o1}
reveals that only five of the forty-one IT \hii regions employed in our
analysis lie in this range and, of these, only three have observed He$^{++}$.
As a result, only $\sim$~3 of the IT \hii regions ($< 10\%$) might have
an $icf > 1$, while still being consistent with the observed values of $\eta$,
[OIII]/[OI], and HeII/H$\beta$.

\section{Matter-Bounded Versus Radiation-Bounded}
\label{mbrb}

As mentioned earlier, since the [OI]$\lambda$6300 emission line is formed in
the recombination region, its intensity depends on the optical depth at the
Lyman limit, $\tau_{LL}$.  As $\tau_{LL}$ increases from 1 (matter-bounded)
to 10$^4$ (radiation-bounded), [OI] increases and the [OIII]/[OI] ratio
decreases by more than 3 orders of magnitude.  Since $\eta$ is a ratio of
ratios, the effect of the Lyman limit optical depth largely cancels and
$\eta$ is relatively insensitive to $\tau_{LL}$.

Images of \hii regions show that they are far from being homogeneous,
and this suggests that different ``sectors" could have different values
of $\tau_{LL}$.  Indeed, several studies indicate that a 
considerable fraction of the ionizing photons escape from many \hii 
regions (\eg Oey \& Kennicutt, 1997; Zurita \etal 2000; Rela\~no \etal 
2001).  It is then possible that for the IT-observed \hii regions
used in determining Y$_{\rm P}$, parts are matter bounded with a very
large -- local -- [OIII]/[OI] ratio, while the remaining radiation
bounded sectors have much lower [OIII]/[OI] ratios.  Thus, while the
observed [OIII]/[OI] ratio might be large, suggesting that no ionization
correction need be made (BFM), this may not necessarily be true.  Indeed,
as has already been seen, for the IT range in [OIII]/[OI], the $icf$
is expected to differ from unity.

To explore the effect of \hii regions with matter-bounded pieces, several
test cases were run in which the \hii region was divided into two sectors,
a radiation-bounded part with [OIII]/[OI] $\ll$ 300, and a matter-bounded
piece with $\tau_{LL} \leq$ 10.  For model \hii regions with matter-bounded 
sectors having $\tau_{LL} \sim$ 10, ratios of [OIII]/[OI] $\geq$ 300 can be 
achieved for covering factors of the radiation-bounded part in the range 0.3 
to 0.8.  As a result, \hii regions might have very large [OIII]/[OI] ratios 
because they are ionized by very massive starbusts (in which case $icf 
\rightarrow 1$), or they may have matter-bounded sections and result from 
much smaller starbursts.  For two-sector models, as the contribution from 
the matter-bounded piece increases, the [OIII]/[OI] ratio increases while 
$\eta$ shifts to slightly higher values (compared to the corresponding 
radiation-bounded model).  In these cases the $icf$, although tending 
towards unity, remains very close to that of the radiation-bounded model 
provided that [OIII]/[OI] $\leq$ 300.  Thus, although partially matter-bounded 
\hii regions might be observed with very high [OIII]/[OI] ratios, suggesting 
an $icf \rightarrow 1$, the observed values of [OIII]/[OI] and $\eta$ provide 
some insurance against being lured into this trap.  Since the [OIII]/[OI] 
line ratio depends on the number of ionizing photons and on the fraction 
of the solid angle in the outer parts of the \hii region which are 
radiation-bounded, it is important that observers estimate the total flux 
in H$\beta$ and the fraction of the flux included in the slit.  This would 
permit a better comparison with the models and, by comparing the observed 
$Q_{H}$ for the whole \hii region and the observed [OIII]/[OI] with the 
model predictions it might be possible to distinguish density-bounded 
\hii regions from those which are radiation-bounded. 

The discussion here, as well as that in \S3, provides a reminder of the
importance of a careful treatment of the recombination transition zone to
accurate, quantitative predictions of the model \hii region $icf - \eta$
relations.  In this connection we note that SJ {\it assume} that the
effects of radiative transfer can be treated in the ``On-The-Spot"
approximation (OTS).  We have tested this assumption and find it to be
unjustified.  For the spectra of our 0.0, 3.3, and 4.5 Myr models, at
fixed $\eta$ the OTS underestimates the $icf$ (the $icf$ is too small)
by a few percent.  In contrast, for the 2.5 Myr starburst the OTS
overestimates the $icf$ by $\ga 0.25$.  For the 5.4 Myr starburst
the differences are even more dramatic.  For this model, when He/H
= 0.083 the $icf - \eta$ relation shown in Fig.~3 is shifted upwards
(the $icf$ is overestimated), similar to the shift for the 2.5 Myr
model, but in this case by $\ga 0.3$.  However, for this same starburst,
when He/H = 0.075 the character of the $icf - \eta$ relation changes
completely from that in Fig.~3 to resemble the one for He/H = 0.083
and, the $icf$ is shifted upwards to $icf > 1.25$.  Thus, although our
results are in excellent {\it qualitative} agreement with those of SJ,
we caution that a careful treatment of radiative transfer is crucial
for {\it quantitatively} accurate, model-predicted \hii region $icf$s.

\section{Discussion}

In outward appearance, as well as in their internal structure, \hii regions
evolve as they age.  In their infancy \hii regions ionized by young, hot,
metal-poor stars have ionized helium present in the zone where hydrogen is
making the transition from ionized to neutral, resulting in an $icf < 1$.
As the starburst ages, the radiation spectrum softens, the $icf$ increases,
and in their early youth \hii regions will have $icf > 1$.  So far, no
detectable amounts of doubly-ionized helium will be present.  As \hii
regions enter middle age, the radiation spectrum hardens and the $icf$
decreases, once again dropping below unity.  By now the spectrum has
enough sufficiently hard photons that detectable amounts of doubly-ionized
helium should be present.  Finally, as \hii regions enter old age, the
spectrum softens, the $icf$ increases ($icf > 1$) one last time, and
they fade away.  But, how does an \hii region show its age?  Is there
a chronometer?  The proposed ``radiation softness" parameters should be
surrogate chronometers but, are they?

As Figure~\ref{etao3o1} reveals, the [OIII]/[OI] ratio is a very poor
chronometer.  Notice that for [OIII]/[OI] $\ga 10$, the {\it same} value
of [OIII]/[OI] corresponds to {\it all} five spectra studied here.  While
the [OIII]/[OI] ratio provides an excellent measure of the {\it intensity}
of the ionizing radiation, it reveals nothing about the age of the \hii
region, or the hardness of its spectrum.  The situation is better for the
Vilchez-Pagel parameter $\eta$.  Figure~\ref{etao3o1} suggests that if
log~$\eta~\ga 0.7$ were observed, then the \hii region would likely be
dominated by a starburst for which $icf > 1$.  Similarly, it is likely
that $icf < 1$ for those \hii regions with log~$\eta~\la -0.1$ (see Fig.~5).
However, starbursts of all ages (except 2.5 Myr) could be responsible for
\hii regions with $-0.1 ~\la$ log~$\eta~\la 0.4$.  Of these, the oldest
starburst has an $icf > 1$, while all the others have an $icf < 1$.
Unfortunately, even in this case (low $\eta$), the Vilchez-Pagel radiation
softness parameter {\it alone} is incapable of distinguishing among starbursts
of $t = $ 0.0, 3.3, and 4.5 Myr, even though the latter two, older starbursts
have very different spectra from the youngest one.

At the same time that Figure~\ref{etao3o1} exposes the inadequacy of
the [OIII]/[OI] ratio, and the deficiencies of $\eta$, it suggests that
$\eta$ and [OIII]/[OI] {\it in combination} might provide a more useful
gauge of \hii region ages by helping to separate the effects of ``hard"
versus ``soft" spectra (which determine whether $icf < 1$ or $> 1$) from
those related to the intensity of the starburst (which determines by {\it
how much} the $icf$ will differ from unity).  In addition, for log~$\eta~\la
0.5$, the presence or absence of He$^{++}$ may help to discriminate between
infancy (no observable He$^{++}$ at $t \approx 0$) and maturity (detectable
He$^{++}$ present for $3~\la t~\la 5$ Myr); see Fig.~\ref{he2vseta}.  For
example, \hii regions with [OIII]/[OI]$~\ga 150$ and $0.3~\la $log~$\eta~\la 
0.5$ are likely ionized by an old ($\sim 5.4$ Myr) starburst and should have
detectable He$^{++}$, while those with the same large values of [OIII]/[OI],
but with $-0.1~\la $log~$\eta~\la 0.2$, should be dominated by a very young
($t \approx 0$) starburst and they should have no observable He$^{++}$.
With reference to Figs.~3, 5, and 6, this approach may be applied to the
IT data.  While many observational constraints are needed for a good 
photoionization model of a specific giant \hii region (SJ), we will account 
for the range of \hii regions included in the IT data set using these 
three parameters.  It is clear that more detailed data for any individual 
\hii region may lead to a better model for that specific region, resulting 
in a slightly different derived value for the $icf$.   However, the agreement 
between our results (below) and those of SJ suggest that these differences 
are likely very small for the bulk of the IT \hii regions.

Of the 41 IT \hii regions in our data set, 36 have log~$\eta~\la 0.25$ 
(see Fig.~5) and are likely ionized by radiation with spectra similar to 
those of our $t = 0.0$, 3.3, or 4.5 Myr starbursts.  For these \hii regions, 
$icf < 1$.  Of the remaining five \hii regions, only three have He$^{++}$ 
detected; only these three are likely to have significant or dominant 
contributions from a spectrum similar to that of our 5.4 Myr starburst 
for which $icf > 1$.  It seems likely then that for more than 90\% (38/41) 
of the IT \hii regions, $icf < 1$ and Y $<$ Y(IT).  This dominance of 
$icf < 1$ is not at all surprising when considering that an \hii region 
spends most of its life ionized by a hard radiation spectrum with 
Q(He$^{0}$)/Q(H$^{0}$) $> 0.15$ (see Fig.~2).  To estimate the magnitude 
of the ionization correction it is necessary to use the available data 
for {\it both} log~$\eta$ {\it and} log([OIII]/[OI]) since for the three 
starbursts with hard spectra, the model-predicted $icf$s vary significantly 
(see Fig.~3) over the range in $\eta$ covered by the IT data 
($-0.3~\la $log~$\eta~\la 0.25$).  However, for log([OIII]/[OI]) $\ga 1.7$
(see Fig.~5), the $icf$s for these three starburst models are nearly 
identical and, for log([OIII]/[OI]) $\la 2.9$, they do differ from unity 
(see Fig.~6).  As a result of this comparison, we conclude that for 
$\sim 38/41$ of the IT \hii regions, $0.95~\la icf~\la 0.99$.  Since 
the IT helium abundance determinations range from Y $\approx 0.24$ to 
Y $\approx 0.26$, this suggests a correction to Y$_{\rm P}$(IT) in the 
range $0.002~\la$$-\Delta$Y$~\la 0.010$, in excellent agreement with the 
SJ estimate of a ~$\sim 2 - 4~\%$ overestimate in the helium abundance, 
corresponding to $0.005~\la$$-\Delta$Y$~\la 0.010$.  Approximating our
correction as $\Delta$Y$~\approx -0.006 \pm 0.002$ and combining it in 
quadrature with the IT primordial helium estimate of Y$_{\rm P}$(IT)$ 
= 0.244 \pm 0.002$, leads to a revised estimate of Y$_{\rm P} = 0.238 
\pm 0.003$.

\section{Conclusions}

The history of the study of the helium-hydrogen $icf$ extends over
several decades (see VGS for references to early work) and the different
model assumptions may have obscured the fact that all such studies reach
the same general conclusions.  \hii regions ionized by massive, intense
starbursts will be large and, when radiation bounded, the transition zone
from ionized to neutral gas is relatively thin compared to the overall
size of the \hii region.  For such \hii regions the $icf \rightarrow 1$.
As the intensity of the starburst diminishes, the relative size of the
transition zone increases and the $icf$ deviates from unity.  Whether the
deviation will be $icf < 1$ or $icf > 1$ is determined by the relative
hardness of the {\it spectrum} of the starburst, {\it not} by its intensity.
Although there will be some quantitative differences among the predicted
$icf$s depending on the details of the spectra adopted (\eg compare VGS, BFM,
and SJ), it is clear that the ratio of helium-ionizing to hydrogen-ionizing
photons is key to determining whether the $icf < 1$ or $> 1$.  Thus, although
the spectra adopted here for starbursts of differing ages differ somewhat
from the corresponding spectra in SJ, we agree with them that there is
a ``critical" value of Q(He$^{0}$)/Q(H$^{0}$) $\approx 0.15$ separating
\hii regions with $icf < 1$ from those with $icf > 1$.  An ideal radiation
softness parameter would provide a unique reflection of this ratio.  The 
Vilchez-Pagel parameter $\eta$, although imperfect, is a valuable surrogate.
However, even when it can be determined that the $icf < 1$ (or, $> 1$), 
the amount by which the $icf$ deviates from unity depends on the intensity
of the radiation spectrum.  For more intense (more massive) starbursts, 
$\eta$ decreases, [OIII]/[OI] increases, and the $icf \rightarrow 1$.  
However, the majority of the \hii regions used to infer the primordial 
helium abundance have values of $\eta$ and [OIII]/[OI] which lie in those
ranges where the models suggest significant (\ie few percent) deviations
of the $icf$ from unity.  Note that for an uncorrected helium mass fraction
in the range Y $\approx$ 0.24 -- 0.26 (IT), a 3\% deviation in the $icf$
from unity corresponds to a change in helium abundance of order 0.006,
three times as large as the IT error on their inferred primordial helium
mass fraction.  Indeed, in our previous analysis we estimated that the IT
value of Y$_{\rm P} = 0.244 \pm 0.002$ needed to be reduced by approximately
0.003 (VGS).  The analysis presented here, in agreement with that of SJ,
strengthens this conclusion and suggests that the reduction may be even
larger.

The IT primordial helium estimate (Y$_{\rm P} = 0.244 \pm 0.002$) is already
somewhat low in comparison to the BBN-predicted value based on the baryon
density inferred from the deuterium observations of O'Meara \etal (2001): for
D/H = $3.0 \pm 0.4 \times 10^{-5}$, $\Omega_{\rm B}h^{2} = 0.020 \pm 0.002$,
and Y$_{\rm P}({\rm BBN; D/H}) = 0.247 \pm 0.001$.  If, instead, the CMB
estimate of the baryon density is used, $\Omega_{\rm B}h^{2} = 0.022
\pm 0.002$ (\eg Kneller \etal 2001 and references therein), the predicted
primordial helium abundance is even larger: Y$_{\rm P}({\rm BBN; CMB})
= 0.248 \pm 0.001$.  Here we have corrected the IT estimate for unobserved
neutral hydrogen and/or helium, concluding that Y$_{\rm P} = 0.238 \pm 0.003$,
thus increasing the tension between the expected and inferred primordial
helium abundances.  Almost certainly, the discussion here will not be the
final word on the primordial abundance of helium.

\vskip 0.5truecm


\noindent {\bf Acknowledgments}

We are pleased to thank the referee, Manuel Peimbert, for valuable 
comments and suggestions.  In Brazil the work of S.M.V. and R.G. is 
partially supported by grants from CNPq (304077/77-1 and 306122/88-0), 
from FAPESP (00/06695-0), and from PRONEX/FINEP (41.96.0908.00); in 
the U.S. the work of G.S. is supported at The Ohio State University by 
DOE grant DE-AC02-76ER-01545.  Some of this work was done while S.M.V. 
was visiting the OSU Physics Department and while G.S. was visiting 
IAGUSP and they wish to thank the respective host institutions for 
hospitality.


\vskip 0.5truecm

\beginapjbib

\bibitem Armour, M.-H., Ballantyne, D.R., Ferland, G.J., Karr, J. \&
Martin, P.G. 1999, PASP, 111, 1251

\bibitem Bahcall, J.N., Pinsonneault, M.H., \& Basu, S. 2001, ApJ,
555, 990

\bibitem Ballantyne, D.R., Ferland, G.J., \& Martin, P.G. 2000,
ApJ, 536, 773 (BFM)

\bibitem Conti, P.S. \& Vacca, W.D. 1994, ApJL, 494, L97

\bibitem Conti, P.S., Leitherer, C., \& Vacca, W.D. 1996, ApJL, 461, L87

\bibitem Cid-Fernandes, R., Dottori, H., Gruenwald, R., \& Viegas, S.M.
1992, MNRAS, 255, 165

\bibitem Gruenwald, R. \& Viegas, S.M. 1992, ApJS, 78, 153 (AANGABA)

\bibitem Gruenwald, R. \& Viegas, S.M. 2000, ApJ, 543, 889

\bibitem Izotov, Y.I., Thuan, T.X., \& Lipovetsky, V.A. 1994
ApJ, 435, 647 (ITL)

\bibitem Izotov, Y.I., Thuan, T.X., \& Lipovetsky, V.A. 1997,
ApJS, 108, 1  (ITL)

\bibitem Izotov, Y.I. \& Thuan, T.X. 1998, ApJ, 500, 188 (IT)

\bibitem Kneller, J.P., Scherrer, R.J., Steigman, G., \& Walker, T.P. 2001,
Phys. Rev. D, In Press (astro-ph/0101386)

\bibitem Oey, M.S. \& Kennicutt, R.C. 1997, MNRAS, 291, 827

\bibitem Olive, K.A., Skillman, E. \& Steigman, G. 1997, ApJ, 483, 788 (OSS)

\bibitem Olive, K.A., \& Steigman, G. 1995, ApJS, 97, 49 (OS)

\bibitem Olive, K.A., Steigman, G., \& Walker, T.P. 2000, Physics Reports,
333-334, 389

\bibitem O'Meara, J.M. \etal 2001, ApJ, 552, 718

\bibitem Pagel, B.E.J., Simonson, E.A., Terlevich, R.J. \& Edmunds, M.
1992, MNRAS, 255, 325

\bibitem Peimbert, M., Peimbert, A., \& Ruiz, M.T. 2000, ApJ, 541, 688

\bibitem Rela\~no, M., Peimbert, M. \& Beckman, J.E. 2001, ApJ, in press 
(astro-ph/0109113)

\bibitem Sauer, D. \& Jedamzik, K. 2001, preprint (astro-ph/0104392) (SJ)

\bibitem Skillman, E.D. 1989, ApJ, 347, 883

\bibitem Stasinska, G. 1990, A\&A Suppl., 83, 501


\bibitem Viegas, S.M., Gruenwald, R., \& Steigman, G. 2000, ApJ, 531, 813 
(VGS)

\bibitem Vilchez, J.M. \& Pagel, B.E.J. 1988, MNRAS, 231, 257

\bibitem Zurita, A., Rozas, M., \& Beckman, J.E. 2000, A\&A, 363, 9

\endapjbib

\end{document}